\documentclass[12pt,preprint]{aastex}

\begin{document}
\title{Is GRB 050904 a super-long burst?}
\author{Y. C. Zou, Z. G. Dai and D. Xu \thanks{Department of Astronomy,
Nanjing University, Nanjing 210093, China; zouyc@nju.edu.cn, dzg@nju.edu.cn.}}
\begin{abstract}
By considering synchrotron radiative process in the internal shock model and
assuming that all internal shocks are nearly equally energetic, we analyze the
gamma-ray burst (GRB) emission at different radii corresponding to different
observed times. We apply this model to GRB 050904 and find that our analytical
results can provide a natural explanation for the multi-band observations of
GRB 050904. This suggests that the X-ray flare emission and the optical
emission of this burst could have originated from internal shocks being due to
collisions among nearly-equally-energetic shells ejected from the central
engine. Thus GRB 050904 appears to be a burst with super-long central engine
activity.

\end{abstract}


\section{Introduction}\label{intro}
The gamma-ray burst (GRB) 050904 was an explosive event at redshift of
$z=6.29\pm 0.01$, which has been measured through various methods
\citep{kawai05, price05, haislip05}. After the Swift trigger
\citep{cummings05}, multi-wavelength observations of this high-redshift burst
were performed. The $\gamma$-ray isotropic-equivalent energy of this burst
$E_{\gamma, iso}$ was between $6.6\times10^{53}\,\rm{ergs}$ and $3.2\times
10^{54}\, \rm{ergs}$ \citep{cusumano05}. \citet{boer05} observed the optical
emission at times of 86 s to 1666 s. The optical afterglow and its spectrum
were analyzed quickly \citep{haislip05, tagliaferri05}. A break of the optical
afterglow light curve was also observed at about $2.6\pm 1.0$ days
\citep{tagliaferri05}. The optical emission appears to be understood in either
both the late internal shock model or the reverse-forward shock model
\citep{wei05}.

One of the most remarkable features of this burst is a long lasting variability
of the X-ray emission, showing several X-ray flares (XRFs). \citet{watson05}
and \citet{cusumano05} suggested that this variability may be due to a
long-lasting activity of the central engine. The XRFs were also found in some
other GRBs \citep{nousek06,obrien06}. \citet{burrows05}, \citet{fan05},
\citet{zhang05} and \citet{wu05} assumed late central engine activities to
interpret the XRFs. The plausible origin models of XRFs were recently proposed,
e.g., the magnetic activity of a newborn millisecond pulsar \citep{dai06},
fragmentation of a neutron star by a black hole in a compact object binary
\citep{faber06}, fragmentation of an accretion disk \citep{perna06}, and
magnetic barrier--driven modulation of an accretion disk \citep{proga06}.
However, there has not yet been any explicit and detailed investigation on this
highly-variable light curve. Enlightened by the internal shock model to
interpret the $\gamma$-ray emission \citep{rees94} and prompt optical emission
\citep{meszaros99}, we assume that all the highly-variable X-ray emission (even
at very late times) originate from internal shocks. These shocks occur when
many faster shells with equal energy and equal mass catch up with a slower
shell with much more kinetic energy and mass. By fitting the peaks of the X-ray
and optical light curves, we find that these internal shocks can lead to the
observed X-ray and optical emission. We describe an internal-shock model in \S
\ref{theory}, fit the peaks of X-ray and optical light curves of GRB 050904 in
\S \ref{GRB050904}. We summarize our results in \S \ref{discussion}.

\section{Model}\label{theory}
\begin{figure}
  \includegraphics[width=1\textwidth]{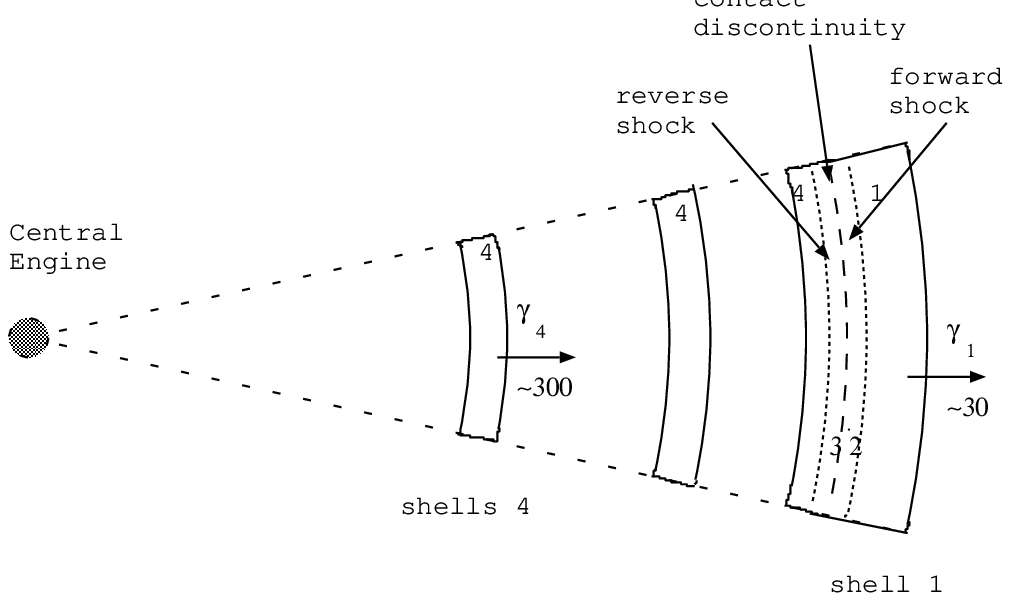}
  \caption{Sketch of our model: A slower shell labeled ``1'' with
  much more energy expands in the front. Many similar faster shells labeled ``4''
  with less energy and less number density catch up with ``shell 1'', and
  reverse-forward shocks occur. The gamma rays and early X-rays are mainly
  emitted from the reverse shocks.}
  \label{fig:sketch}
\end{figure}

We assume two cold shells labeled with 1 and 4, whose isotropic kinetic
energies are $E_1$ and $E_4$, widths are $\Delta_{1,0}$ and $\Delta_{4,0}$ in
the cosmological rest frame, and Lorentz factors are $\gamma_1$ and $\gamma_4$
($\gamma_4 \gg \gamma_1 \gg 1$), respectively. The shells come from the central
engine with time lag $T$ (also in the rest frame). Shell 4 catches up with
shell 1 at radius $r_{\rm{int}}\simeq 2\gamma_1^2 c T$ and then two internal
shocks occur, that is, a forward shock that propagates into shell 1 and a
reverse shock that propagates into shell 4. Four regions are divided by the
reverse and forward shocks: unshocked shell 1 (region 1), shocked gas of shell
1 (region 2), shocked gas of shell 4 (region 3) and unshocked shell 4 (region
4), with a contact discontinuity (CD) surface separating region 2 and region 3.
Figure \ref{fig:sketch} gives a sketch of this model. The number densities of
region 1 and region 4 at radius $r$ are $n_1=E_1/(4\pi r^2 \gamma_1^2
\Delta_{1,0} m_p c^2) $ and $n_4=E_4/(4\pi r^2 \gamma_4^2 \Delta_{4,0} m_p
c^2)$ respectively. To produce plenty of X-rays and $\gamma$-rays, it is
required that the reverse-shocked materials are relativistic, i.e.,
$\gamma_{34} \simeq \frac{1}{\sqrt{2}}(\frac{\gamma_4}{\gamma_1})^{1/2}f^{-1/4}
\gg 1$ \citep{sari95, wu03, zou05}, where $\gamma_{34}$ is the relative Lorentz
factor between regions 4 and 3, and the number density ratio is defined as
$f\equiv {n_4}/{n_1}$.

To calculate the width of the coasting shell, we define the spreading radius
$R_\Delta \equiv \gamma^2 \Delta_0$ \citep{meszaros93, piran93}. For the
internal shocks of our interests, if $r < \min(R_{\Delta,1} = 1.0\times 10^{15}
\gamma_{1,1.5}^2 \Delta_{1,0,12} \mbox{cm}, R_{\Delta,4} = 1.0\times 10^{16}
\gamma_{4,3}^2 \Delta_{4,0,10} \mbox{cm})$, the width of the shells can be
considered to be constant, and for  $r > \max(R_{\Delta,1}, R_{\Delta,4})$, the
width should be regarded as $r/\gamma_1^2$ and $r/\gamma_4^2$ for the spreading
of the shells. The conventional notation $Q_k=Q/10^k$ with cgs units is used in
this paper except for special explanations. We now consider these two cases.

\subsection{No-spreading case}\label{nospreading}
At radius $r<\min(R_{\Delta,1}, R_{\Delta,4})$, for a certain set of parameters
$\Delta_{1,0,12}=\Delta_{4,0,10}=E_{1,54}=E_{4,51}= \gamma_{1,1.5}=
\gamma_{4,3}=1$, one obtains $\gamma_{14}^2 f \simeq 2.5 \times 10^{-2}
E_{1,54}^{-1} E_{4,51} \Delta_{1,0,12} \Delta_{4,0,10}^{-1} \ll 1$ (where
$\gamma_{14}$ is the relative Lorentz factor between regions 4 and 1), implying
that the forward shock is Newtonian, and  $\gamma_{14}^2/f \simeq 2.5\times
10^{6}  E_{1,54} E_{4,51}^{-1} \Delta_{1,0,12}^{-1} \Delta_{4,0,10}
\gamma_{1,1.5}^{-4} \gamma_{4,3}^4 \gg 1$, which shows that the reverse shock
is relativistic. Using the shock conditions \citep{blandford76}, and keeping
equality of pressures and Lorentz factors of the materials at two sides of the
contact discontinuity \citep{sari95}, we can obtain the number densities $n_2,
n_3$, and energy densities $e_2, e_3$ of the shocked regions in the comoving
frame,
\begin{eqnarray}
  n_2 &\simeq& 1.0 \times 10^{10}\,(1+z)^2\,E_{1,54}\,\Delta_{1,0,12}^ {- 1 }
  \,\gamma_{1,1.5}^ {- 6 }\,t_{\oplus,2}^ {- 2}
  \,\mbox{cm}^{-3},\\
  n_3 &\simeq& 9.3 \times 10^{6} (1+z)^2 \,E_{4,51}\,\Delta_{4,0,10}^ {- 1 }
  \,\gamma_{1,1.5}^ {- 5 }\,\gamma_{4,3}^ {- 1 }\,t_{\oplus,2}^ {- 2 }
  \,\mbox{cm}^{-3},\\
  e_2 &=& e_3 \simeq 2.2 \times 10^{5} (1+z)^2 \,E_{4,51}\,\Delta_{4,0,10}
  ^ {- 1 }\,\gamma_{1,1.5}^ {- 6 }\,t_{\oplus,2}^ {- 2 }
  \,\mbox{erg cm}^{-3}.
\end{eqnarray}
The Lorentz factor $\gamma_2 = \gamma_3 \simeq \gamma_1 \simeq 31.6
\gamma_{1,1.5}$. Here $t_\oplus$ is the observer's time.

Assuming that the reverse shock lasts from $r_{int}$ to $r_{int}+\delta r$, we
find $\delta r \simeq 1.0\times 10^{13}\gamma_{1,1.5}^2
\Delta_{4,0,10}\mbox{cm}$, which corresponds to a pulse from the beginning to
peak. In the observer's frame ($dt_{\oplus} = (1+z)dr/(2\gamma^2c)$), this
lasting time is
\begin{equation}
  \delta t_{\oplus} \simeq 0.17 (1+z) \Delta_{4,0,10} \,\mbox{s}.
\end{equation}

By considering the synchrotron emission at the reverse shock-crossing time,
which corresponds to the peak emission time, we obtain the characteristic
frequencies $\nu_{m,2,p}$ and $\nu_{m,3,p}$, cooling frequencies $\nu_{c,2,p}$
and $\nu_{c,3,p}$, synchrotron self-absorption frequencies $\nu_{a,2,p}$ and
$\nu_{a,3,p}$, and the maximum flux densities $f_{\nu,\max,2,p}$ and
$f_{\nu,\max,3,p}$ in the observer's frame \citep[see][for original formulae of
synchrotron emission]{zou05}
\begin{eqnarray}
  \nu_{m,2,p} &\simeq&
  1.3 \times 10^{11}\,E_{1,54}^ {- 2 }\,E_{4,51}^{{{5
      }\over{2}}}\,\Delta_{1,0,12}^2\,
  \Delta_{4,0,10}^ {- {{5}\over{2}} }\,\epsilon_{B,-1}^{{{1}\over{2}}}
  \,\epsilon_{e,-0.5}^2\,\gamma_{1,1.5}^ {- 2 }\,\zeta_{1/6}^2
  \,t_{\oplus,2}^ {- 1 } \, \mbox{Hz},
  \\
  \nu_{m,3,p} &\simeq& 1.5 \times 10^{17}\,\,E_{4,51}^{{{1}\over{2}}}
  \,\Delta_{4,0,10}^ {- {{1}\over{2}} }
  \,\epsilon_{B,-1}^{{{1}\over{2}}}\,\epsilon_{e,-0.5}^2\,\gamma_{1,1.5}
  ^ {- 4 }\,\gamma_{4,3}^2\,\zeta_{1/6}^2 \,t_{\oplus,2}^ {- 1 } \, \mbox{Hz},
  \\
  \nu_{c,2,p} &=& \nu_{c,3,p} \simeq 1.2 \times 10^{15}\,(1+z)^ {- 4 }
  \,E_{4,51}^ {- {{3}\over{2}} }
  \,\Delta_{4,0,10}^ {- {{1}\over{2}} }\,
  \epsilon_{B,-1}^ {- {{3}\over{2}} }\,\gamma_{1,1.5}^8 \,t_{\oplus,2}^3  \,\mbox{Hz},
  \\
  \nu_{a,2,p} &\simeq& 3.2 \times 10^{14}\,(1+z)^ {- {{5}\over{18}} }
  \,E_{1,54}^ {- {{7}\over{36}} }\, E_{4,51}^{{{5}\over{9}}}
  \,\Delta_{1,0,12}^{{{7}\over{36}}}\,\Delta_{4,0,10}^ {- {{5}\over{9}} }
  \,\epsilon_{B,-1}^{{{1}\over{12}} }\,\epsilon_{e,-0.5}^{{{1}\over{3}}}
  \,\gamma_{1,1.5}^ {- {{8}\over{9}} }\,\zeta_{1/6}^{{{1}\over{3}}}
  \,t_{\oplus,2}^ {- {{13}\over{18}} }\, \mbox{Hz},
  \\
  \nu_{a,3,p} &\simeq& 1.2 \times 10^{14}\,(1+z) ^ {- {{1}\over{3}} }
  \,E_{4,51}^{{{1}\over{3}}} \,\Delta_{4,0,10}^ {- {{1
      }\over{3}} }\,\gamma_{1,1.5}^ {- {{2}\over{3}} }\,\gamma_{4,3}^ {- {{1
      }\over{3}} }\,t_{\oplus,2}^ {- {{2}\over{3}} } \, \mbox{Hz},
  \\
  f_{\nu,\max,2,p} &\simeq&
  8.0 \times 10^{2}\,D_{28}^ {- 2 }\,E_{1,54}^{{{1
      }\over{2}}}\,E_{4,51}\,(1+z)^2\,
  \Delta_{1,0,12}^ {- {{1}\over{2}} }\,\epsilon_{B,-1}^{{{1}\over{2}}}
  \,\gamma_{1,1.5}^ {- 2 }\,t_{\oplus,2}^ {- 1 } \,\mbox{Jy},
  \\
  f_{\nu,\max,3,p}&\simeq& 0.47 \,(1+z)^2
  \,D_{28}^ {- 2 }\,E_{4,51}^{{{3
      }\over{2}}}\,\Delta_{4,0,10}^ {- {{1
      }\over{2}} }\,\epsilon_{B,-1}^{{{1}\over{2}}}\,\gamma_{1,1.5}^ {- 2 }
  \,\gamma_{4,3}^ {- 1 } \,t_{\oplus,2}^ {- 1 } \,\mbox{Jy},
\end{eqnarray}
where $\zeta_{1/6}=6(p-2)/(p-1)$, $p$ is the spectral index of the
shock-accelerated electrons, $D$ is the luminosity distance from the burst to
the observer (as a function of redshift $z$), and $\epsilon_e$ and $\epsilon_B$
are the fractions of the internal energy density that were carried by electrons
and magnetic fields, respectively. In this paper, $p$ is taken to be $2.2$, and
$\epsilon_e$ and $\epsilon_B$ are equal for both region 2 and region 3. Note
that the expressions of $\nu_{a,2,p}$ and $\nu_{a,3,p}$ are only valid in the
cases of $\nu_{c,2,p}<\nu_{m,2,p}<\nu_{a,2,p}$ and
$\nu_{a,3,p}<\nu_{c,3,p}<\nu_{m,3,p}$.

From the above equations, we see $\nu_{c,2,p} < \nu_{m,2,p} < \nu_{a,2,p}$ and
$\nu_{a,3,p} < \nu_{c,3,p} < \nu_{m,3,p}$ for typical parameters. The flux
density in the fast-cooling case at the peak time is then given by
\[
f_{\nu,2,p} \simeq
\]
\begin{eqnarray}
{\small
  \left\{
  \begin{array}{ll}
    2.4 \times 10^{8}\,(1+z)^ {- 1 }\,D_{28}^ {- 2 }
    \,E_{4,51}^ {- 1 }\, \epsilon_{B,-1}^ {- 1 }\,\gamma_{1,1.5}^8
    \,\nu_{18}^2\,\, t_{\oplus,2}^4
    \,\mbox{Jy},
    & \nu < \nu_{c,2,p} < \nu_{m,2,p} < \nu_{a,2,p},
    \\
    7.1 \times 10^{9}\,(1+z)\,D_{28}^ {- 2 }\,E_{4,51}^ {- {{1}\over{4}} }
    \,\Delta_{4,0,10}^{{{1}\over{4}}}\, \epsilon_{B,-1}^ {- {{1}\over{4}} }
    \,\gamma_{1,1.5}^4\,\nu_{18}^{{{5}\over{2}} }\, t_{\oplus,2}^{{{5}\over{2}}}
    \,\mbox{Jy},
    & \nu_{c,2,p} < \nu  < \nu_{a,2,p} ,
    \\
    2.0 \times 10^{-3}\,D_{28}^ {- 2 }
    \,E_{1,54}^ {- {{ 7}\over{10}} }\,E_{4,51}^{{{7}\over{4}}}
    \,\Delta_{1,0,12}^{{{7}\over{10}}}\,\Delta_{4,0,10} ^ {- {{7}\over{4}} }
    \,\epsilon_{B,-1}^{{{1}\over{20}}}\, \epsilon_{e,-0.5}^{{{6}\over{5}}}
    \,\gamma_{1,1.5}^{{{4}\over{5}}}\, \nu_{18}^ {- {{11}\over{10}} }
    \,\zeta_{1/6}^{{{6}\over{5}}}\,t_{\oplus,2}^ {- {{1 }\over{10}} }
    \,\mbox{Jy},
    & \nu_{c,2,p} < \nu_{m,2,p} < \nu_{a,2,p} < \nu.
    \\
  \end{array}
  \right.
  \label{eq:ff3_nu_2_nospreading}
}
\end{eqnarray}
for the forward shock emission, and
\[
f_{\nu,3,p} \simeq
\]
\begin{eqnarray}
  \left\{
  \begin{array}{ll}
    3.3 \times 10^{8}\,(1+z)^{-1}\,D_{28}^ {- 2 }\,E_{4,51}^ {- 1
    }\,\epsilon_{B,-1}^ {- 1 }\,\gamma_{1,1.5}^8
    \,\nu_{18}^2 \,t_{\oplus,2}^4  \,\mbox{Jy},
    & \nu<\nu_{c,3,p}<\nu_{a,3,p}<\nu_{m,3,p},
    \label{eq:f_nu_nos_nuCAM}
    \\
    9.6 \times 10^{9}\,(1+z)^{{10}\over{3}}
    \,D_{28}^ {- 2 }\,E_{4,51}^ {- {{1}\over{4}} }\,
    \Delta_{4,0,10}^{{{1}\over{4}}}\,\epsilon_{B,-1}^ {- {{1}\over{4}} }
    \,\gamma_{1,1.5}^4\,\nu_{18}^{{{5}\over{2}}}\,t_{\oplus,2}^{{{5}\over{2}}}
    \, \mbox{Jy},
    & \nu_{c,3,p}<\nu<\nu_{a,3,p}<\nu_{m,3,p},
    \label{eq:f_nu_nos_CnuAM}
    \\
    1.6 \times 10^{-2}\,\,D_{28}^ {- 2 }\,E_{4,51}^{{{3
    }\over{4}}}\,
    \Delta_{4,0,10}^ {- {{3}\over{4}} }\,\epsilon_{B,-1}^ {- {{1}\over{4
    }} }\,\gamma_{1,1.5}^2\,\gamma_{4,3}^ {- 1 }\,\nu_{18}^ {- {{1}\over{2}}}
    \,t_{\oplus,2}^{{{1}\over{2}}} \,\mbox{Jy},
    & \nu_{c,3,p}<\nu_{a,3,p}<\nu<\nu_{m,3,p},
    \label{eq:f_nu_nos_CAnuM}
    \\
    5.2 \times 10^{-3}\,\,D_{28}^ {- 2 }\,E_{4,51}^{{{21
    }\over{20}}}\,\Delta_{4,0,10}^ {- {{21}\over{20}} }\,
    \epsilon_{B,-1}^{{{1}\over{20}}}\,\epsilon_{e,-0.5}^{{{6}\over{5}}}
    \,\gamma_{1,1.5}^ {- {{2}\over{5}} }\,\gamma_{4,3}^{{{1}\over{5}}}\,
    \,\zeta_{1/6}^{{{6}\over{5}}} \nu_{18}^ {- {{11}\over{10}} }
    \,t_{\oplus,2}^ {- {{1}\over{10}} } \,\mbox{Jy},
    & \nu_{c,3,p}<\nu_{a,3,p}<\nu_{m,3,p}<\nu,
    \label{eq:f_nu_nos_CAMnu}
  \end{array}
  \right.
  \label{eq:ff_nu_3_nospreading}
\end{eqnarray}
for the reverse shock emission.

\subsection{Spreading case}\label{spreading}
In the case of $r > \max(R_{\Delta,1}, R_{\Delta,4})$, the widths of the shells
become $\Delta_1 \simeq r/\gamma_1^2$ and $\Delta_4 \simeq r/\gamma_4^2$
because of the spreading effect. For the same parameters as
\S\ref{nospreading}, $\gamma_{14}^2 f \simeq 0.5 E_{1,54}^{-1} E_{4,51}
\gamma_{1,1.5}^{-2} \gamma_{4,3}^2 < 1$, showing that the forward shock can be
considered as Newtonian approximately, and $\gamma_{14}^2/f \simeq 2.5\times
10^4 E_{1,54} E_{4,51}^{-1} \gamma_{1,1.5}^{-2} \gamma_{4,3}^2 \gg 1$, implying
that the reverse shock is still relativistic.

Using the shock conditions as in \S\ref{nospreading}, we obtain
\begin{eqnarray}
  n_2 &\simeq& 1.7 \times 10^{9}\,(1+z)^3\,E_{1,54}
  \,\gamma_{1,1.5}^ {- 6 }\,t_{\oplus,2}^ {- 3 } \,\mbox{cm}^{-3},
  \label{eq:n2_spreading}
  \\
  n_3 &\simeq& 1.5 \times 10^{7} (1+z)^3 \,E_{4,51}
  \,\gamma_{1,1.5}^ {- {7} }\,\gamma_{4,3}
  \, t_{\oplus,2}^ {- 3 } \,\mbox{cm}^{-3},
  \label{eq:n3_spreading}
  \\
  e_2 &=& e_3 \simeq 3.7 \times 10^5 (1+z)^3  \,E_{4,51}
  \,\gamma_{1,1.5}^ {- 8 }\,\gamma_{4,3}^ { 2 }\,t_{\oplus,2}^ {- 3}\,\mbox{erg cm}^{-3},
  \\
  \gamma_2 &=& \gamma_3 \simeq \gamma_1 \simeq 31.6 \gamma_{1,1.5}.
\end{eqnarray}
The distance for the reverse shock to cross shell 4 is $\delta r \simeq 6.0 \times
10^{12} (1+z)^{-1} \gamma_{4,3}^{-1} \gamma_{1,1.5}^2 t_{\oplus,2} \rm{cm}$,
and the rising time of the pulse in the observer's frame is $\delta t_{\oplus}
\simeq 0.2 \gamma_{4,3}^{-1}  t_{\oplus,2} \rm{s}$.

The corresponding emission values at the peak time are given by
\begin{eqnarray}
  \nu_{m,2,p} &\simeq& 1.6\times 10^{13}\,(1+z)^{{{1}\over{2}}}
  \,E_{1,54}^ {- 2 }\,E_{4,51}^{{{5}\over{2}}}
  \,\epsilon_{B,-1}^{{{1}\over{2}}}\,\epsilon_{e,-0.5}^2\,\gamma_{1,1.5}^ {- 7 }
  \,\gamma_{4,3}^5\,\zeta_{1/6}^2\,t_{\oplus,2}^ {- {{3}\over{2}} }
  \,\mbox{Hz},
  \\
  \nu_{m,3,p} &\simeq& 2.0 \times 10^{17} \,(1+z)^ {{1}\over{2}}
  E_{4,51}^ { {{1}\over{2}} } \,\epsilon_{B,-1}^{{{1}\over{2}}} \,\epsilon_{e,-0.5}^2\,
  \gamma_{1,1.5}^ {- {{5}\over{2}} }\,\gamma_{4,3}^{3}\,
  \zeta_{1/6}^2 \,t_{\oplus,2}^ {- {{3}\over{2}} }
  \,\mbox{Hz},
  \\
  \nu_{c,2,p} &=& \nu_{c,3,p} \simeq 1.5 \times 10^{15} \,(1+z)^ {-{{7}\over{2}}}
  \,E_{4,51}^{-{{3}\over{2}}}
  \,\epsilon_{B,-1}^ {- {{3}\over{2}} } \,\gamma_{1,1.5}^{7}
  \,\gamma_{4,3} \,t_{\oplus,2}^{{{5}\over{2}} }
  \,\mbox{Hz},
  \\
  \nu_{a,2,p} &\simeq& 6.4 \times 10^{14}\,(1+z)^{{{1}\over{12}}}
  \,E_{1,54}^ {- {{7}\over{36}} }\, E_{4,51}^{{{5}\over{9}}}
  \,\epsilon_{B,-1}^{{{1}\over{12}}}
  \,\epsilon_{e,-0.5}^{{{1}\over{3}}}
  \,\gamma_{1,1.5}^ {- 2 }\,\gamma_{4,3}^{{{10}\over{9}}}
  \,\zeta_{1/6}^{{{1}\over{3}}}\,t_{\oplus,2}^ {-{{13}\over{12}} }
  \,\mbox{Hz},
  \\
  \nu_{a,3,p} &\simeq& 2.0 \times 10^{13} \,(1+z)^ {{14}\over{5}}
  \,E_{4,51}^ {{{9}\over{5}} }
  \,\epsilon_{B,-1}^{{{6}\over{5}}} \,\gamma_{1,1.5}^ {- 8 }
  \,\gamma_{4,3}^ {- {{1}\over{5}} }\,t_{\oplus,2}^ {- {{19}\over{5}} }
  \,\mbox{Hz},
  \\
  f_{\nu,\max,2,p} &\simeq&  0.033 \,(1+z)^{{{5}\over{2}}}
  \,D_{28}^ {- 2 }\,E_{1,54}^{{{1}\over{2}}}\,E_{4,51}
  \,\epsilon_{B,-1}^{{{1}\over{2}}}\,\gamma_{1,1.5}^ {- 2 }
  \,t_{\oplus,2}^ {- {{3}\over{2}} }
  \,\mbox{Jy},
  \\
  f_{\nu,\max,3,p}&\simeq& 0.6 \,(1+z)^{{5}\over{2}} \,D_{28}^ {- 2 }
  \,E_{4,51}^{{{3}\over{2}}}
  \,\epsilon_{B,-1}^{{{1}\over{2}}}
  \,\gamma_{1,1.5}^ {- 3}
  \,t_{\oplus,2}^ {- {{3}\over{2}} }
  \,\mbox{Jy}.
\end{eqnarray}
Note that the expressions of $\nu_{a,2,p}$ and $\nu_{a,3,p}$ are valid only in
the cases of $\nu_{m,2,p}<\nu_{c,2,p}<\nu_{a,2,p}$ and
$\nu_{a,3,p}<\nu_{c,3,p}<\nu_{m,3,p}$ respectively.

In the case of $\nu_{m,2,p}<\nu_{c,2,p}<\nu_{a,2,p}$, the corresponding flux
density of the forward shock emission is
\[
f_{\nu,2,p} \simeq
\]
\begin{eqnarray}
{\small
  \left\{
  \begin{array}{ll}
    2.1 \times 10^{7}\,(1+z)\,D_{28}^ {- 2 }\,E_{1,54}^ {- 1 }
    \,E_{4,51}\,\epsilon_{e,-0.5}\,\gamma_{1,1.5}\,\gamma_{4,3}^2
    \,\nu_{18}^2\,\zeta_{1/6}\,t_{\oplus,2}^2
    \,\mbox{Jy},
    & \nu < \nu_{m,2,p} < \nu_{c,2,p} < \nu_{a,2,p} ,
    \\
    5.2 \times 10^{9}\,(1+z)^{{{ 3}\over{4}}}\,D_{28}^ {- 2 }
    \,E_{4,51}^ {- {{1}\over{4}} }\,\epsilon_{B,-1}^ {- {{1 }\over{4}} }
    \,\gamma_{1,1.5}^{{{9}\over{2}}}\,\gamma_{4,3}^ {- {{1 }\over{2}} }
    \,\nu_{18}^{{{5}\over{2}}}\,t_{\oplus,2}^{{{11}\over{4}}}
    \,\mbox{Jy},
    & \nu_{m,2,p} < \nu < \nu_{a,2,p},
    \\
    1.7 \times 10^{-2}\,(1+z)^{{{21}\over{20}}}\,D_{28}^ {- 2 }
    \,E_{1,54}^ {- {{ 7}\over{10}} }\,E_{4,51}^{{{7}\over{4}}}
    \,\epsilon_{B,-1}^{{{1}\over{20}}} \,\epsilon_{e,-0.5}^{{{6}\over{5}}}
    \,\gamma_{1,1.5}^ {- {{27}\over{10 }} }\,\gamma_{4,3}^{{{7}\over{2}}}
    \,\nu_{18}^ {- {{11}\over{10}} }\, \zeta_{1/6}^{{{6}\over{5}}}
    \, t_{\oplus,2}^ {- {{23}\over{20}} }
    \,\mbox{Jy},
    & \nu_{m,2,p} < \nu_{c,2,p} < \nu_{a,2,p} < \nu .
  \end{array}
  \right.
  \label{eq:fs1_nu_2_spreading}
}
\end{eqnarray}
As the cooling frequency $\nu_{c,2,p}$ exceeds the self-absorption frequency
$\nu_{a,2,p}$ at time $t_{\oplus}\simeq 70 (1+z)  E_{1,54}^{-7/134}
E_{4,51}^{37/67} \Delta_{1,0,12}^{7/134} \Delta_{4,0,10}^{-1/67}
\epsilon_{B,-1}^{57/134} \epsilon_{e,-0.5}^{6/67} \gamma_{1,1.5}^{-160/67}
\zeta_{1/6}^{6/67}\,$s, the corresponding emission after this time is then
\[
f_{\nu,2,p} \simeq
\]
\begin{eqnarray}
{\small
  \left\{
  \begin{array}{ll}
    2.9 \times 10^{7}\,(1+z)\,D_{28}^ {- 2 }\,E_{1,54}^ {- 1  }
    \,E_{4,51}\,\epsilon_{e,-0.5}\,\gamma_{1,1.5}\,\gamma_{4,3}^2
    \,\nu_{18}^2\,\zeta_{1/6}\,t_{\oplus,2}^2
    \,\mbox{Jy},
    & \nu < \nu_{m,2,p} < \nu_{a,2,p} < \nu_{c,2,p},
    \\
    7.1 \times 10^{9}\,(1+z)^{{{3}\over{4}}}\,D_{28}^ {- 2 }
    \,E_{4,51}^ {- {{ 1}\over{4}} }\,\epsilon_{B,-1}^ {- {{1}\over{4}} }
    \,\gamma_{1,1.5}^{{{9}\over{2}}}\, \gamma_{4,3}^ {- {{1}\over{2}} }
    \,\nu_{18}^{{{5}\over{2}}}\,t_{\oplus,2}^{{{11}\over{4}}}
    \,\mbox{Jy},
    & \nu_{m,2,p} < \nu < \nu_{a,2,p} < \nu_{c,2,p} ,
    \\
    4.4 \times 10^{-1}\,(1+z)^{{{14}\over{5}}}\,D_{28}^ {- 2 }
    \,E_{1,54}^ {- {{ 7}\over{10}} }\,E_{4,51}^{{{5}\over{2}}}
    \,\epsilon_{B,-1}^{{{4}\over{5}}}\, \epsilon_{e,-0.5}^{{{6}\over{5}}}
    \,\gamma_{1,1.5}^ {- {{31}\over{5}} } \,\gamma_{4,3}^3
    \,\nu_{18}^ {- {{3}\over{5}} }\,\zeta_{1/6}^{{{6 }\over{5}}}
    \, t_{\oplus,2}^ {- {{12}\over{5}} }
    \,\mbox{Jy},
    & \nu_{m,2,p} < \nu_{a,2,p} < \nu < \nu_{c,2,p} ,
    \\
    1.7 \times 10^{-2}\,(1+z)^{{{21}\over{20}}}\,D_{28}^ {- 2 }
    \,E_{1,54}^ {- {{ 7}\over{10}} }\,E_{4,51}^{{{7}\over{4}}}
    \,\epsilon_{B,-1}^{{{1}\over{20}}} \,\epsilon_{e,-0.5}^{{{6}\over{5}}}
    \,\gamma_{1,1.5}^ {- {{27}\over{10}} }\,\gamma_{4,3}^{{{7}\over{2}}}
    \,\nu_{18}^ {- {{11}\over{10}} }\, \zeta_{1/6}^{{{6}\over{5}}}
    \, t_{\oplus,2}^ {- {{23}\over{20}} }
    \,\mbox{Jy},
    & \nu_{m,2,p} < \nu_{a,2,p} < \nu_{c,2,p} < \nu .
  \end{array}
  \right.
}
  \label{eq:fs2_nu_2_spreading}
\end{eqnarray}

The flux density of the reverse shock emission in the case of
$\nu_{a,3,p}<\nu_{c,3,p}<\nu_{m,3,p}$ is given by
\[
f_{\nu,3,p} \simeq
\]
\begin{eqnarray}
{\small
  \left\{
  \begin{array}{ll}
    3.5 \times 10^{8} \,(1+z)^{-1} \,D_{28}^ {- 2 }
    \,E_{4,51}^{-1} \,\epsilon_{B,-1}^ {- 1 }
    \,\gamma_{1,1.5}^8 \,\nu_{18}^2 \,t_{\oplus,2}^4
    \,\mbox{Jy},    & \nu<\nu_{a,3,p}<\nu_{c,3,p}<\nu_{m,3,p},
    \label{eq:f_nu_s_nuACM}
    \\
    5.3 \,(1+z)^{{{11}\over{3}}} \,D_{28}^ {- 2 }
    \,E_{4,51}^2
    \,\epsilon_{B,-1} \,\gamma_{1,1.5}^ {- {{16}\over{3}} }
    \,\gamma_{4,3}^ {-{{1}\over{ 3}} } \,\nu_{18}^{{{1}\over{3}}}
    \,t_{\oplus,2}^ {- {{7}\over{3}} }
    \,\mbox{Jy},    & \nu_{a,3,p}<\nu<\nu_{c,3,p}<\nu_{m,3,p},
    \label{eq:f_nu_s_AnuCM}
    \\
    2.3 \times 10^{-2} \,(1+z)^{{{3}\over{4}}} \,D_{28}^ {- 2 }
    \,E_{4,51}^{{{3}\over{4}}}
    \,\epsilon_{B,-1}^ {- {{1}\over{4}}}
    \,\gamma_{1,1.5}^{{{1}\over{2}}} \,\gamma_{4,3}^{{{1}\over{2}}}
    \,\nu_{18}^ {- {{1}\over{2}} } \,t_{\oplus,2}^ {- {{1}\over{4}} }
    \,\mbox{Jy},    & \nu_{a,3,p}<\nu_{c,3,p}<\nu<\nu_{m,3,p},
    \label{eq:f_nu_s_ACnuM}
    \\
    8.9 \times 10^{-3} \,(1+z)^ {{{21}\over{20}} } \,D_{28}^ {- 2 }
    \,E_{4,51}^{{{21}\over{20}}}
    \,\epsilon_{B,-1}^{{{1}\over{20}}}
    \,\epsilon_{e,-0.5}^{{{6}\over{5}}} \,\gamma_{1,1.5}^ {- {{5}\over{2}} }
    \,\gamma_{4,3}^{{{23}\over{10}}} \,\zeta_{1/6}^{{{6}\over{5}}}
    \,\nu_{18}^ {- {{11}\over{10}} }\,t_{\oplus,2}^ {- {{23}\over{20}} }
    \,\mbox{Jy},    & \nu_{a,3,p}<\nu_{c,3,p}<\nu_{m,3,p}<\nu.
    \label{eq:f_nu_s_ACMnu}
  \end{array}
  \right.
  \label{eq:ff_nu_3_spreading}
}
\end{eqnarray}
At time $t_{\oplus} \simeq 3.4 \times 10^2 (1+z) E_{4,51}^{1/2}
\epsilon_{B,-1}^{1/2} \epsilon_{e,-0.5}^{1/2} \gamma_{1,1.5}^{-19/8}
\gamma_{4,3}^{1/2} \zeta_{1/6}^{1/2}$s, $\nu_{c,3,p} = \nu_{m,3,p}$. After that
time, the order of the break frequencies becomes
$\nu_{a,3,p}<\nu_{m,3,p}<\nu_{c,3,p}$. In this case, the flux density becomes
\[
f_{\nu,3,p} \simeq
\]
\begin{eqnarray}
  \label{eq:fs_nu_3_spreading}
  \left\{
  \begin{array}{ll}
    3.4 \times 10^{9} \,(1+z) \,D_{28}^ {- 2 }
    \,\epsilon_{e,-0.5}
    \,\gamma_{1,1.5}^2 \,\gamma_{4,3} \,\zeta_{1/6} \,\nu_{18}^2 \,t_{\oplus,2}^2
    \,\mbox{Jy},    & \nu<\nu_{a,3,p}<\nu_{c,3,p}<\nu_{m,3,p},
    \label{eq:f_nu_s_nuAMC}
    \\
    1.0 \,(1+z)^{{{7}\over{3}}} \,D_{28}^ {- 2 }
    \,E_{4,51}^{{4}\over{3}}
    \,\epsilon_{B,-1}^{{1}\over{3}} \,\epsilon_{e,-0.5}^{-{{2}\over{3}}}
    \,\gamma_{1,1.5}^ {- {{4}\over{3}} }
    \,\gamma_{4,3}^ {-{1} } \,\zeta_{1/6}^{-{{2}\over{3}}} \,\nu_{18}^{{{1}\over{3}}}
    \,t_{\oplus,2}^ {- 1 }
    \,\mbox{Jy},    & \nu_{a,3,p}<\nu<\nu_{c,3,p}<\nu_{m,3,p},
    \label{eq:f_nu_s_AnuMC}
    \\
    2.3 \times 10^{-1} \,(1+z)^{{{14}\over{5}}} \,D_{28}^ {- 2 }
    \,E_{4,51}^{{{9}\over{5}}}
    \,\epsilon_{B,-1}^ {{{4}\over{5}}} \,\epsilon_{e,-0.5}^{{6}\over{5}}
    \,\gamma_{1,1.5}^{-6} \,\gamma_{4,3}^{{{9}\over{5}}}
    \,\zeta_{1/6}^{{6}\over{5}}
    \,\nu_{18}^ {- {{3}\over{5}} } \,t_{\oplus,2}^ {- {{12}\over{5}} }
    \,\mbox{Jy},    & \nu_{a,3,p}<\nu_{c,3,p}<\nu<\nu_{m,3,p},
    \label{eq:f_nu_s_AMnuC}
    \\
    8.9 \times 10^{-3} \,(1+z)^ {{{21}\over{20}} } \,D_{28}^ {- 2 }
    \,E_{4,51}^{{{21}\over{20}}}
    \,\epsilon_{B,-1}^{{{1}\over{20}}}
    \,\epsilon_{e,-0.5}^{{{6}\over{5}}} \,\gamma_{1,1.5}^ {- {{5}\over{2}} }
    \,\gamma_{4,3}^{{{23}\over{10}}} \,\nu_{18}^ {- {{11}\over{10}} }
    \,\zeta_{1/6}^{{{6}\over{5}}} \,t_{\oplus,2}^ {- {{23}\over{20}} }
    \,\mbox{Jy},    & \nu_{a,3,p}<\nu_{c,3,p}<\nu_{m,3,p}<\nu.
    \label{eq:f_nu_s_AMCnu}
  \end{array}
  \right.
\end{eqnarray}
From equations (\ref{eq:ff_nu_3_spreading}) and (\ref{eq:fs_nu_3_spreading}),
we can see that the flux densities are the same for the case in which the
observed frequency exceeds the three break frequencies.

We consider that many shells like the foregoing shell 4 collide with shell 1 at
different radii $r_{int}$, where $r_{int}$ indicates the observed time by the
relation $t_{\oplus} \simeq (1+z)r/(2\gamma_3^2 c)$. Thus the peaks of emission
obey a power law temporal profile. Under more realistic conditions, ``shell 1''
may be sole, proceeding at the front, and many shells like ``shell 4'' ejected
from the central engine continually overtake ``shell 1'' (see Figure
\ref{fig:sketch}). As the energy of ``shell 1'' is assumed to be much greater
than that of ``shell 4'', the dynamics of ``shell 1'' after the collision can
be almost unchanged, and thus the above relations are still applicable. In the
popular collapsar model, ``shell 1'' can be understood as due to the envelope
of a collapsing massive star. Because a relativistic outflow needs to
accumulate enough energy to break out of the envelope and is inevitably
polluted by the baryons in the envelope, this outflow (i.e., ``shell 1") should
be more energetic but have less Lorentz factor comparing with ``shell 4''.

\section{GRB 050904}\label{GRB050904}
\begin{figure}
  \includegraphics[width=1.2\textwidth]{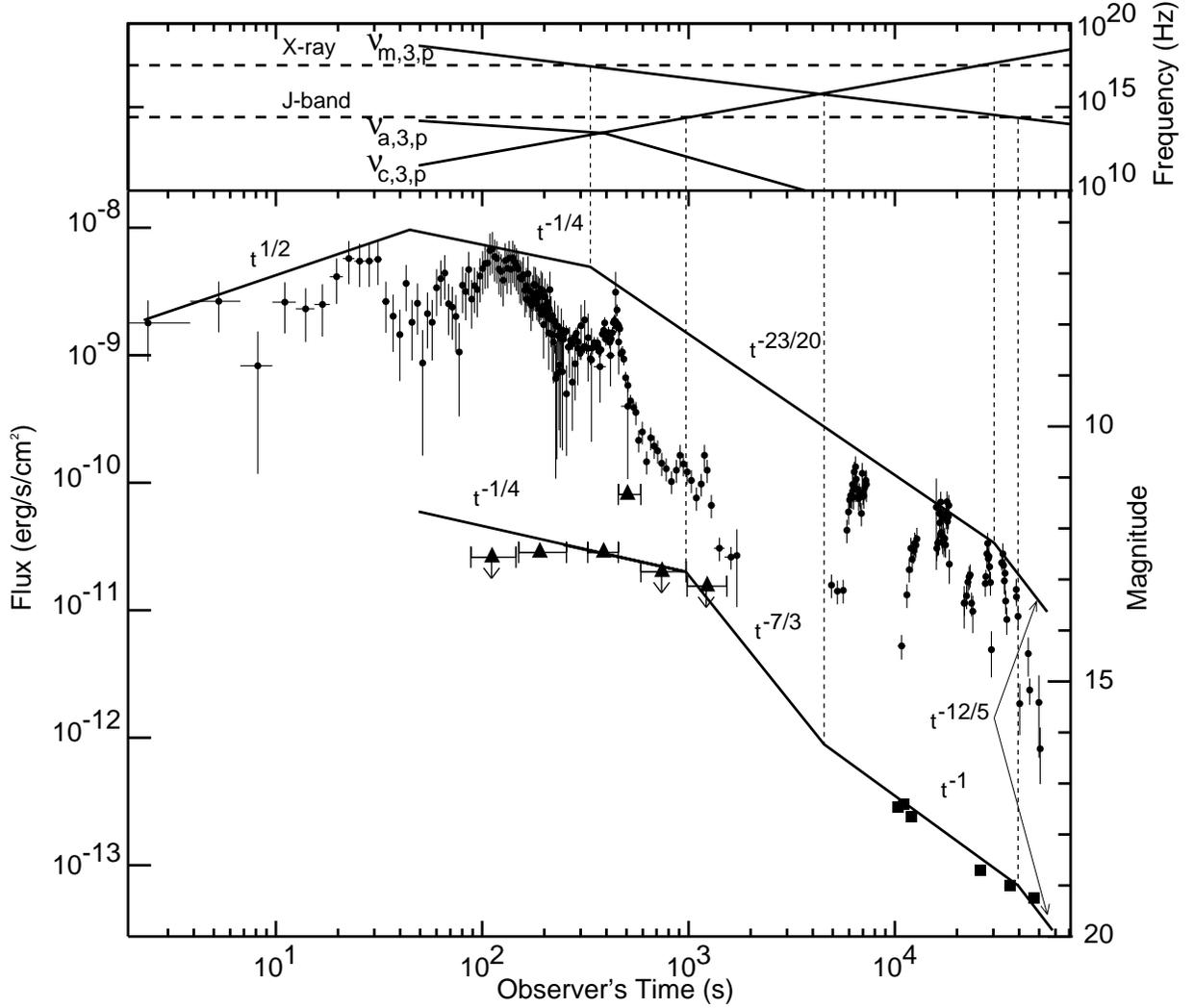}
  \caption{Fitting to the peaks of X-ray and optical light curves of GRB 050904
  (solid line). The observed X-ray data (points with error bars) are taken
  from \citet{cusumano05} where the flux is divided by $(1+z)^2$ and the time
  is multiplied by $(1+z)$ to give the observer's frame. The optical data
  are taken from \citet{boer05} (triangles with error bars, extrapolated to
  J-band by the relation $F_\nu \propto \nu^{1/3}$, which is the synchrotron
  emission in the case $\nu_a<\nu<\nu_c<\nu_m$) and \citet{haislip05} (squares).
  The top panel shows the break frequencies as functions of time. Vertical dashed
  lines represent the crossing times of any two frequencies.}
  \label{fig:X-peaks}
\end{figure}

Using the above model, we fit the peaks of the X-ray light curves of GRB 050904
in the observer's frame. We consider the model parameters $E_1=10^{55}{\rm erg},
E_4=1.1\times 10^{51}{\rm erg}, \gamma_1=31.6, \gamma_4=1455, \Delta_{1,0}=
10^{12}{\rm cm}, \Delta_{4,0}=2.0\times 10^8 {\rm cm}, \epsilon_B=0.1,
\epsilon_e=0.316$, and $p=2.2$ and the cosmology with $\Omega_m=0.23,
\Omega_\lambda=0.73$ and  $H_0=71 \,{\rm km}\,{\rm s}^{-1}\,{\rm Mpc}^{-1}$. These
parameters lead to a Newtonian forward shock and a relativistic reverse shock.
As shown in Figure \ref{fig:X-peaks}, the solid line with four segments is
fitted analytically. The flux is an integral of flux density in a frequency
range of $4\times 10^{16}$Hz to $2\times 10^{18}$Hz, which corresponds to XRT's
0.2-10 keV band. The temporal breaks are due to crossing of the three break
frequencies.

The first segment represents the emission from a relativistic reverse shock
with a constant shell width, which satisfies equation (\ref{eq:f_nu_nos_CAnuM})
with temporal index $1/2$. At time $\sim 42$ s, the corresponding radius is
$R(42\,{\rm s}) \simeq 2 \gamma_3^2 c t_\oplus/(1+z) \simeq 3.5\times 10^{14}\,
\rm{cm}$, while the spreading radii of region 1 and region 4 are $R_{s,1}
\simeq \gamma_1^2 \Delta_{1,0} \simeq 10^{15}\,\rm{cm}$, and $R_{s,4} \simeq
4.2\times 10^{14}\,\rm{cm}$ respectively. The three radii are approximately
equal, and thus $R(42\,{\rm s})$ can be taken as the common spreading radius of
regions 1 and 4.

After the first break time, regions 1 and 4 should be spreading and the flux
density is described by the third expression of equation
(\ref{eq:f_nu_s_ACnuM}). The temporal index is $-1/4$. Up to $\sim 350\,$s, the
frequency $\nu_{m,3,p} \simeq 2.6 \times 10^{17}$Hz decreases to the observed
frequencies of XRT, and then the order of the break frequencies becomes
$\nu_{a,3,p}<\nu_{c,3,p}<\nu_{m,3,p}<\nu$ or
$\nu_{a,3,p}<\nu_{m,3,p}<\nu_{c,3,p}<\nu$. The flux density is given by the
last expression of equation (\ref{eq:f_nu_s_ACMnu}), and the temporal index of
flux is $-23/20$. After $\sim 3\times 10^{4}\,$s, the order is
$\nu_{a,3,p}<\nu_{m,3,p}<\nu<\nu_{c,3,p}$, and the temporal index becomes
$-12/5$.

\begin{figure}
  \includegraphics[width=0.8\textwidth,angle=270]{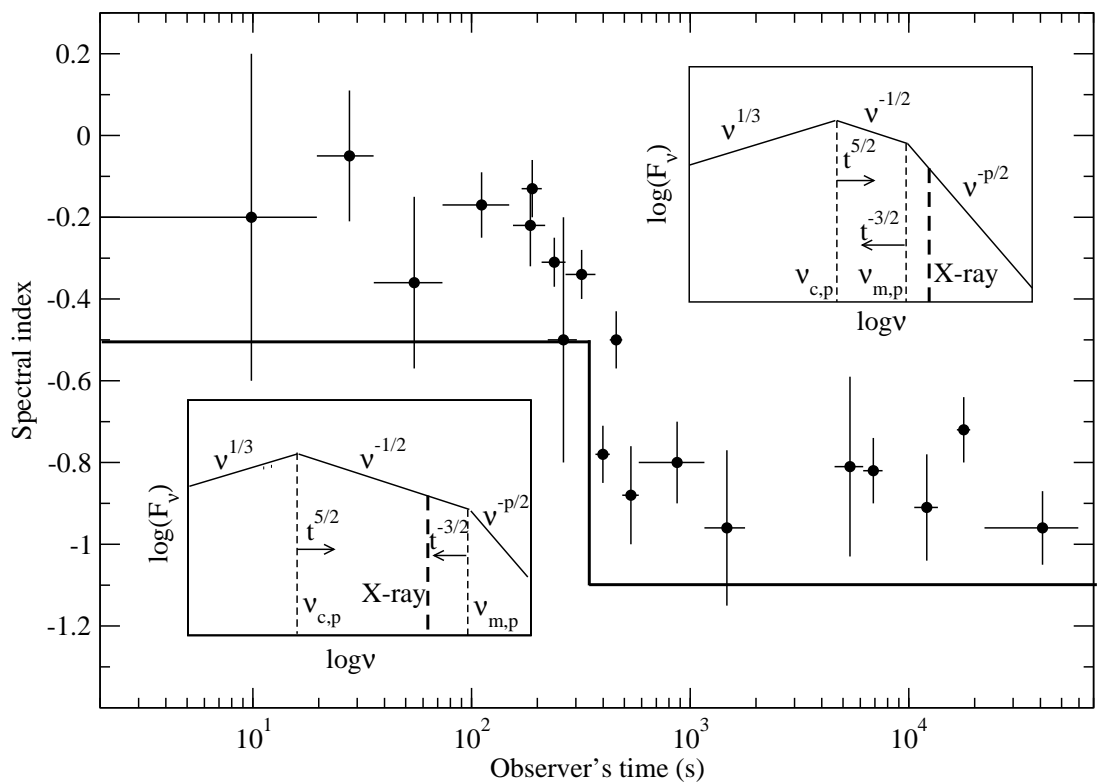}
  \caption{Spectral index evolution. The thick solid line is the expected spectral
  index at the peak of synchrotron radiation. The corresponding spectral
  energy distribution for the two index values ($-0.5$ and $-1.1$) are plotted
  schematically in the two sub-figures. The points with error bars are
  taken from \citet{cusumano05}.}
  \label{fig:spectra}
\end{figure}

Observationally it was reported that the photon indices are $\sim -1.3$ and
$\sim -1.9$ (whose corresponding spectral indices are $-0.3$ and $-0.9$
respectively) at early and late times respectively \citep{cummings05,
watson05}. The observer's time at which the spectral index changed is about
350\,s. This time can be naturally understood as the crossing time of $\nu_m$
to $\nu$. Before this crossing time, $\nu_c<\nu<\nu_m$ and the spectral index
is $-1/2$, and after this crossing time, $\nu_c<\nu_m<\nu$ and the index is
$-1.1$. Figure \ref{fig:spectra} shows the spectral index evolution and a
sketch for the evolution of spectral energy distribution. These spectral
indices are generally slightly less than the observed ones, because our model
only gives the spectral indices at the peak times but the observations were
performed in relatively broad ranges of time. At the beginning of each pair of
reverse and forward shocks, the cooling frequency $\nu_c$ is very large
\citep{zou05}, because this frequency is proportional to $t_{\rm co}^{-2}$,
where $t_{\rm co}$ is the comoving time of the shocks. Thus $\nu_c$ can be
larger than $\nu$ during the initial stage of the shocks. In the case of
$\nu<(\nu_m,\nu_c)$, the spectral index is $1/3$, which makes the observed
spectral index larger than the expected values $-0.5$ and $-1.1$.

\begin{figure}
  \includegraphics[width=1.2\textwidth]{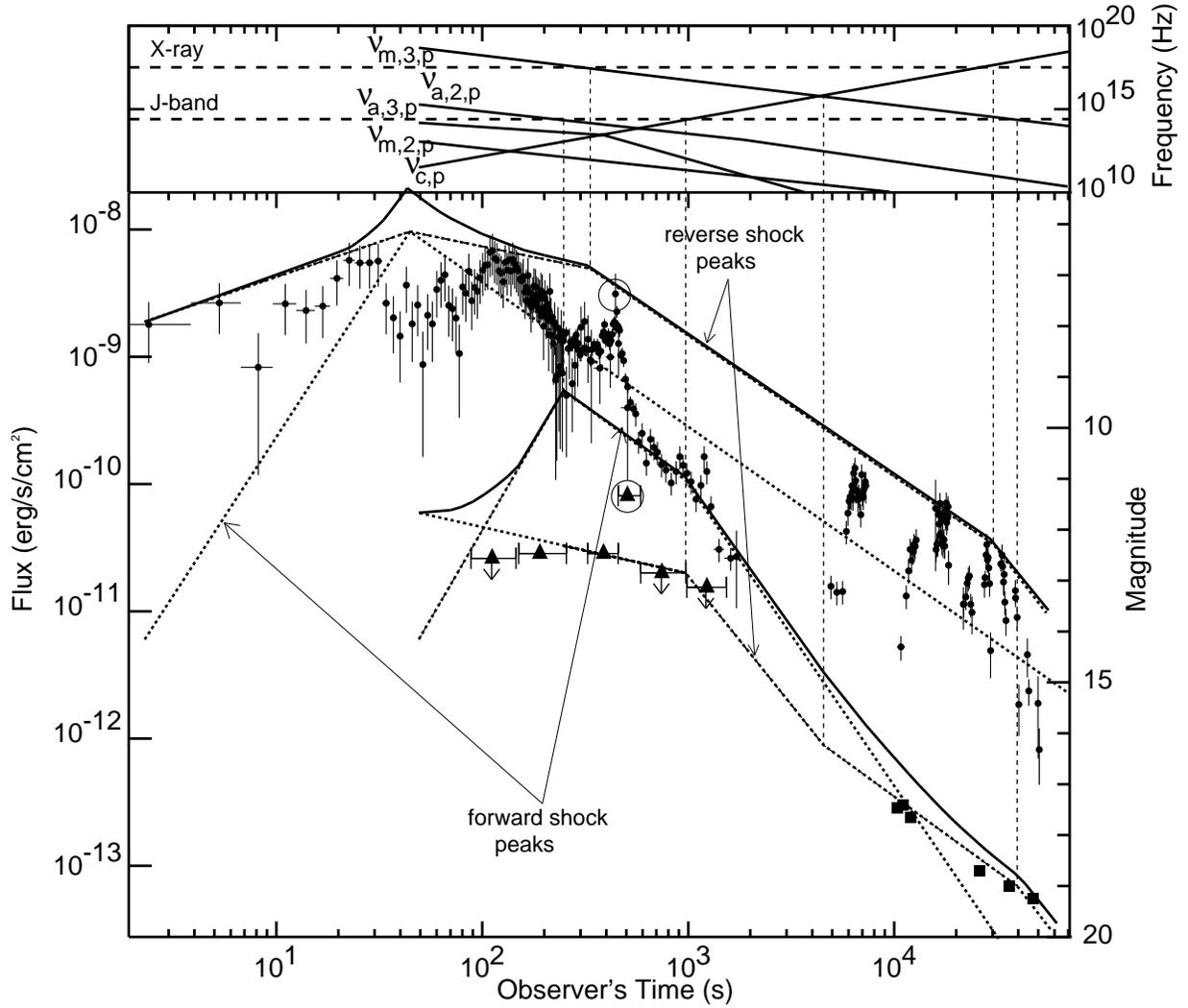}
  \caption{Same as Figure \ref{fig:X-peaks} except for counting in the contribution
  of the forward shocks. The solid line indicates the total peak flux density and
  the others are annotated in this figure.}
  \label{fig:rfs}
\end{figure}

The optical data during the same period were reported \citep{haislip05,
boer05}. The observed J-band ($2.47\times 10^{14}$Hz) data are plotted in
Figure \ref{fig:X-peaks} in units of magnitude. Using the above parameters, we
plot a line representing the J-band peak flux density. However, the J-band
emission may be mainly contributed by the forward shocks. Since the forward
shocks are Newtonian, the emission contributes more at optical band than does
at the X-ray band. Furthermore, as the number density of the forward-shocked
region is much larger than that of the reverse shocks (see equations
(\ref{eq:n2_spreading}) and (\ref{eq:n3_spreading})), the contribution to the
J-band emission exceeds that of the reverse shocks. To show the contribution of
the forward shocks, Figure \ref{fig:rfs} duplicates Figure \ref{fig:X-peaks}
but counts in the forward shocks. It is reasonable to see that the fitted peaks
are greater than the observed data. As the fitting lines stand for the peak
fluxes of the emission of internal shocks, these fluxes should be greater than
the time-averaged observed data especially for the optical emission, which
lasts a relatively longer time for the observation. If there are not
reverse-forward shocks at a certain time, the observed flux density should be
much less than the expected peak one. Therefore, the fitting solid line is
basically consistent with the observed data.

For the highest (fourth) J-band point (449\,s-589\,s) and the X-ray peak nearly
at the same time, we suggest that they could arise from the same internal
shocks, and the X-ray peak could occur only at time slightly earlier than $\sim
440\,$s (we circled them in Figure \ref{fig:rfs} for emphasis). There is a time
lag ($\sim 0-140$s) between the X-ray emission and the J-band emission. This
may be caused by a spectral time lag \citep{norris00}, the different
refractivities of photons in different energy bands during the propagation from
the source to the earth, or a quantum gravity effect (which is also due to
different refractivities in vacuum for different photons, \citet{amelino98}).
There has also been an example with a long time lag, GRB 980425 \citep{norris00}, 
which shows the probability of time lag with tens of seconds.

We further show that our model is consistent with the observed data at some
other aspects. Firstly, the Lorentz factor of region 1 should not change
significantly during the period of X-ray emission. Because faster ejected
shells (region 4) have much less kinetic energy than the front slower one
(region 1) and the forward shocks are Newtonian, the Lorentz factor and the
particle number of region 1 do not increase significantly when it is overtaken
by faster shells. On the other hand, we estimate the deceleration radius $r_d
\equiv [3E_0/(4\pi\gamma^2nm_p c^2)]^{1/3}$, where $E_0$ is the
isotropic-equivalent kinetic energy, and $n$ is the medium density. Letting the
gamma-ray isotropic equivalent energy be $E_{\gamma,iso}=2.5\times 10^{54}$erg,
which is consistent with the observation range \citep{cusumano05}, and assuming
the radiative efficiency $\eta=0.2$ and the medium density $n=3.16
\rm{cm}^{-3}$, we obtain $E_0 = E_{\gamma, iso} (1-\eta)/\eta = 1\times 10^{55}
\rm{erg}$. This energy is consistent with the parameters assumed above. And
then the deceleration radius is $r_d \simeq 8.0 \times 10^{17} E_{0,55}^{1/3}
\gamma_{1.5}^{-2/3} n_{0.5}^{-1/3} \rm{cm}$, which corresponds to the observed
time $t_{\oplus} \simeq 9.7 \times 10^{4}[(1+z)/7.29] E_{0,55}^{1/3}
\gamma_{1.5}^{-8/3} n_{0.5}^{-1/3} \rm{s}$. This deceleration time is larger
than the X-ray duration time, indicating that region 1 is not decelerated by
the ambient medium during this period.

Secondly, the Lorentz factor at the jet-break time \citep{tagliaferri05} is
about $1/\theta_j \simeq 17.5$ ( $\theta_j \simeq 0.057$). It should be greater
at an earlier time. This is in agreement with our used parameter ($\gamma_{1}
=31.6$).


Thirdly, we note the photon indices at the second break of the fitting X-ray
peak line in Figure \ref{fig:X-peaks}. The indices are $-1.22\pm 0.10$ during
the time period of (149s, 223s), $-1.5\pm 0.3$ during the time period of (223s,
303s) in the BAT data, and $-1.13\pm 0.07$ during the time period of (169s,
209s), $-1.31\pm 0.06$ during the time period of (209s, 269s), $-1.34\pm 0.06$
during the time period of (269s, 371s) in the XRT data \citep[Table 1 of][where
the time is divided by $(1+z)$ to give the local observer's frame]{cusumano05}.
It can be seen that the photon indices of the BAT data are greater than those
of the XRT data. In our model, the frequency $\nu_m$ is crossing the observed
frequency during this period. For the frequency orders $\nu_c<\nu<\nu_m$ and
$\nu_c<\nu_m<\nu$, the flux densities are scaled as $f_\nu \propto \nu^{-1/2}$
and $f_\nu \propto \nu^{-11/10}$ respectively. Therefore, the spectrum is
harder for the lower-frequency emission than for the higher-frequency emission,
which is consistent with the observed data.

Fourthly, the mean intrinsic column density is $N_{\rm H} = (2.30\pm 0.50)
\times 10^{22} {\rm cm}^{-2}$, as measured by XRT \citep{cusumano05}. This high
density cannot be explained by the ambient medium with assumed number density
$3 {\rm cm}^{-3}$. Because the radiation originates from the reverse-forward
shocks and should pass through ``shell 1'', we find that the column density of
``shell 1'' is $N_{\rm H} = E_1/(\gamma_1 4\pi r^2 m_p c^2)$, which is $1.67
\times 10^{23} {\rm cm}^{-2}$ at radius $r=10^{16}$cm and $1.67 \times 10^{21}
{\rm cm}^{-2}$ at radius $r=10^{17}$cm. These values are basically consistent
with the measured mean column density. Therefore, the materials in ``shell 1''
provide an explanation for the high column density measured by XRT.

Finally, there are several low peaks in figure \ref{fig:X-peaks} at about
$10^3$s. They may be caused by non-uniform Lorentz factors or energies of the
ejected rapid shells.

In short, the X-ray flares are understood as due to the internal shocks. These
shocks are caused by the continually ejected faster shells that overtake the
sole slower one with much larger kinetic energy and much more baryons.

\section{Conclusions and discussion}\label{discussion}
In this paper, we have proposed a simple internal shock model: a slower massive
shell propagates in the front, and many faster shells with similar mass and
energy catch up with the front slower shell at different radii. These
collisions lead to reverse-forward shocks. The shocks produce the $\gamma$-ray
and lower-energy emission by synchrotron radiation. We obtained a ``normal''
set of parameters to fit the peak fluxes of XRFs of GRB 050904, and found that
our model is well consistent with the observations including the early optical
data.

For GRB 050904, we also found that the contribution of the forward shock
emission is insignificant in the X-ray band, but significant in the J-band at
early times, because the frequency of emission from the Newtonian forward
shocks mainly lies in the optical band. However, we should note that,
for other bursts, forward shocks may also be important for the X-ray emission.
If the leading shell (``shell 1'') is not more energetic than the followed
faster shells (``shells 4''), the velocity of ``shell 1'' cannot be regarded as
unchanged after some collisions and the simple model in section \ref{theory} is
invalid. If the parameters differ, the deceleration radius may be smaller, and
the external shock may become important. All the effects would change the
profile of the XRF peaks.

In general, we argued that some long bursts should have the same properties:
the peaks are nearly constant before the time at which $\nu_m$ is just equal to
the observed frequency $\nu$, and decline rapidly with temporal index about
$-(3p-2)/4$ after this time. Therefore, the $\gamma$-ray emission of these
bursts persists up to the time when $\nu_m=\nu$, and disappears because of the
rapid decline. In this case, the $\gamma$-ray emission ceases when the peak
energy $E_{p}\simeq h\nu$. There are two other reasons for the disappearance of
$\gamma$-rays. One is that the flux of $\gamma$-ray emission decreases below
the detectable limits of telescopes when the frequency order is still
$\nu_c<\nu<\nu_m$. The other is the turnoff of the central engine when $\nu_m$
is still larger than $\nu$. Only in the latter case, it can be said that the
$\gamma$-rays duration is the burst duration.

\acknowledgments We would like to thank the referees for valuable comments that
have allowed us to improve our manuscript and X. F. Wu, L. Shao, Y. F. Huang,
B. Zhang and Y. Z. Fan for helpful discussions. The formulae were generated by
the help of Maxima. This work was supported by the National Natural Science
Foundation of China (grants 10233010 and 10221001).

\clearpage

\clearpage

\clearpage

\clearpage

\end{document}